\pdfoutput=1
\RequirePackage{ifpdf}
\ifpdf % We are running pdfTeX in pdf mode
\documentclass[pdftex]{sigma}
\else
\documentclass{sigma}
\fi

\begin{document}

\allowdisplaybreaks

\renewcommand{\thefootnote}{$\star$}

\renewcommand{\PaperNumber}{069}

\FirstPageHeading

\ShortArticleName{Complex SUSY Transformations and the Painlev\'e IV Equation}

\ArticleName{Complex SUSY Transformations\\ and the Painlev\'e IV Equation\footnote{This
paper is a contribution to the Special Issue ``Superintegrability, Exact Solvability, and Special Functions''. The full collection is available at \href{http://www.emis.de/journals/SIGMA/SESSF2012.html}{http://www.emis.de/journals/SIGMA/SESSF2012.html}}}

\Author{David BERM\'UDEZ}
\AuthorNameForHeading{D.~Berm\'udez}

\Address{Departamento de F\'isica, Cinvestav, AP 14-740, 07000 M\'exico DF, Mexico}
\Email{\href{dbermudez@fis.cinvestav.mx}{dbermudez@fis.cinvestav.mx}}

\ArticleDates{Received July 29, 2012, in f\/inal form September 28, 2012; Published online October 11, 2012}

\Abstract{In this paper we will explicitly work out the complex f\/irst-order SUSY transformation for the harmonic oscillator in order to obtain both real and complex new exactly-solvable potentials. Furthermore, we will show that this systems lead us to exact complex solutions of the Painlev\'e IV equation with complex parameters. We present some concrete examples of such solutions.}


\Classification{81Q60; 35G20; 34M55}

\renewcommand{\thefootnote}{\arabic{footnote}}
\setcounter{footnote}{0}

\vspace{-2mm}

\section{Introduction}

Supersymmetry was introduced in quantum f\/ield theory in order to relate the bosonic and fermionic sectors into one unif\/ied superf\/ield~\cite{WB92}. This theoretical framework is described by some deformations of Lie algebras and their commutation relations. Although there is no expe\-rimental evidence of supersymmetry in nature yet, its concepts have aroused new ideas in other f\/ields of science.

The supersymmetric quantum mechanics (SUSY QM) is one of these new ideas: while f\/irst developed by Witten~\cite{Wit81}, it was soon related to the Darboux transformation and to the facto\-ri\-za\-tion method, which was thought to be completed in 1951 by Infeld and Hull~\cite{IH51} but it was developed further as a paradigm to obtain new exactly-solvable by Mielnik~\cite{Mie84} and others~\cite{BS95,BS97,Fer84,Nie84,Suk85a} and deepen into the algebraic structure of quantum systems.
Even more, for those potentials that are not exactly-solvable we still have a powerful new tool to develop approximation methods, giving rise to a whole new branch of study called \textit{spectral design}~\cite{AFHNNS04}, since one can obtain an exactly-solvable potential with a prescribed spectrum, i.e.\ with some given energy levels.

In this paper we will use the scheme of SUSY QM to obtain complex partner potentials. As far as we know, they were f\/irst developed in~\cite{AICD98}, where complex potentials with real energy spectra were obtained using complex transformation functions associated with real factorization energies and later in~\cite{FMR03} with \textit{complex factorization energies}. On the other hand, in this work we will study complex potentials with \textit{complex energy spectra}.

The theoretical framework of SUSY QM have connections with other branches of study, for example with non-classical orthogonal polynomials and the special functions \cite{SO95}, which are solutions of linear dif\/ferential equations. In this paper, we are interested in some other functions that play an analogous role as special functions, but now for non-linear dif\/ferential equations. Specif\/ically, we are talking about the Painlev\'e equations. Some specialists (for example~\cite{CM08, IKSY91}) consider that solutions of the Painlev\'e equations will be the future members of special functions in the twenty-f\/irst century. It turns out that one of these equations, the Painlev\'e IV equation~($P_{\rm IV}$) (see~\cite{GLS02}), is closely related to potentials obtained by SUSY QM from the harmonic oscillator.
Note that $P_{\rm IV}$ also appears in other areas of physics, for example in f\/luid mecha\-nics~\cite{Win92}, non-linear optics~\cite{FG90}, and quantum gravity~\cite{FIK91}. It also arises in mathematics, e.g., in the symmetric reduction of several partial dif\/ferential equations like Boussinesq equation~\cite{CK89}, dispersive wave equation~\cite{PW89}, non-linear cubic Schr\"odinger equation \cite{BP80}, and auto-dual Yang--Mills f\/ield equations~\cite{AC91}.

The SUSY QM technique has been used before to obtain solutions of $P_{\rm IV}$; however, they are \mbox{either} real or complex but associated with a dif\/ferential equation having real parameters~\mbox{\cite{BF11a,BF12a, BF11b}}. In this work we will expand the solution set to obtain complex solutions, associated now with complex parameters of~$P_{\rm IV}$.

To accomplish this, the structure of the paper is the following: in  Section~\ref{section2} we will work out the complex scheme of the f\/irst-order SUSY QM. In Section~\ref{section3} we will describe the second-order polynomial Heisenberg algebras and its relationship with $P_{\rm IV}$. Then, in Section~\ref{section4} we will use the f\/irst-order SUSY partners of the harmonic oscillator in order to connect with complex solutions to $P_{\rm IV}$. These solutions shall be studied in Section~\ref{section5}. Finally, we will present our conclusions in Section~\ref{section6}.

\section{Complex SUSY QM}\label{section2}
%\label{complex}
Let us begin with a given Hamiltonian $H$ which has been completely solved, i.e., all of its eigenvalues $E_k$ and eigenfunctions $\psi_k$,  $k=0,1,\dots$ are known:
\begin{gather}
H =-\partial^2 +V(x),\label{defH}\\
H\psi_k(x) =E_k\psi_k(x),\label{eigen}
\end{gather}
where $\partial=d/dx$ and we are using natural units such that $\hbar=m=1$.

Now, we use the factorization method which is equivalent to SUSY QM but much simpler in this case. We propose that $H$ is factorizable as
\begin{gather}\label{facH}
H=A^{-}A^{+}+\epsilon.
\end{gather}
In the standard real factorization method there is an extra condition $A^{+}\equiv (A^{-})^{\dag}$. In this paper we do not use this constrain, but rather we simply ask that
\begin{gather}\label{As}
A^{+} =-\partial +\beta(x),\qquad
A^{-} =\partial +\beta(x),
\end{gather}
where $\beta(x)$ is a complex function to be found. This choice represents a more general factorization than the usual real one~\cite{ROM2003}.

Working out the operations in~\eqref{facH}, using the def\/initions in~\eqref{defH} and~\eqref{As}, we obtain one condition for~$\beta(x)$
\begin{gather*}
\beta'+\beta^{2}=V(x)-\epsilon,
\end{gather*}
which is a Riccati equation.

On the other hand, if we consider a similar factorization but in a reversed order and introduce a new Hamiltonian $\tilde{H}$, def\/ined by $\tilde{H}=-\partial^2 +\tilde{V}$ and
\begin{gather}\label{facH2}
\tilde{H}=A^{+}A^{-}+\epsilon,
\end{gather}
it turns out that
\begin{gather*}
\tilde{V}(x)=V(x)-2\beta'(x).
\end{gather*}

Besides, from equations \eqref{facH} and \eqref{facH2} it is straightforward to show that~\cite{AICD98,ROM2003}
\begin{gather}\label{entre1}
\tilde{H}A^{+} =A^{+}H,\qquad
HA^{-} =A^{-}\tilde{H},
\end{gather}
which are the well known \textit{intertwining relationships} with~$A^{+}$,~$A^{-}$ being the \textit{intertwining ope\-ra\-tors}. From equations~\eqref{eigen} and~\eqref{entre1} we can obtain the eigenvalues and eigenfunctions of the new Hamiltonian $\tilde{H}$ as follows
\begin{gather*}
\tilde{H}A^{+}\psi_k(x) =A^{+}H\psi_k(x)=E_{k}A^{+}\psi_k(x),\qquad
\tilde{H}\left[A^{+}\psi_k(x)\right] =E_{k}\left[A^{+}\psi_k(x)\right].
\end{gather*}
Therefore, the eigenfunctions $\tilde{\psi}_k$ of $\tilde{H}$ associated with the eigenvalues $E_k$ become
\begin{gather*}
\tilde{\psi}_k\propto A^{+}\psi_k(x)\propto \frac{W(u,\psi_k)}{u},
\end{gather*}
where $W$ is the Wronskian, $\beta(x)=[\ln\, u(x)]'$, and $u(x)$ is an eigenfunction of $H$ (in general a~non-physical one) associated with a~complex eigenvalue $\epsilon$, i.e.
\begin{gather*}
Hu=\epsilon u.
\end{gather*}
Furthermore, the eigenstates $\tilde{\psi}_k$ are not automatically normalized as in the real SUSY QM since now
\begin{gather*}
\langle A^{+}\psi_n | A^{+}\psi_n \rangle = \langle \psi_n | (A^{+})^{\dag}A^{+} \psi_n \rangle,
\end{gather*}
and in this case $(A^{+})^{\dag}A^{+}\neq (H-\epsilon)$. Nevertheless, since they are normalizable we can introduce a normalizing constant $C_n$, chosen for simplicity as $C_n \in \mathbb{R}^{+}$, so that
\begin{gather*}
\tilde{\psi}_n(x)=C_n A^{+}\psi_n(x),\qquad \langle \tilde{\psi}_n|\tilde{\psi}_n\rangle=1.
\end{gather*}
Finally, there is a wavefunction
\begin{gather*}
\tilde{\psi}_{\epsilon}\propto \frac{1}{u},
\end{gather*}
that is also eigenfunction of $\tilde{H}$
\begin{gather*}
\tilde{H}\tilde{\psi}_{\epsilon}=\epsilon\tilde{\psi}_{\epsilon}.
\end{gather*}
If it is normalizable, it turns out that $\tilde{V}(x)$ is a complex potential which Hamiltonian~$\tilde{H}$ has the following spectrum
\begin{gather}
\text{Sp}(\tilde{H})=\{\epsilon\}\cup\{E_n,n=0,1,\dots\},\label{spectrum}
\end{gather}
and $\epsilon\in\mathbb{C}$ although in particular $\epsilon$ could be real.

\section[Polynomial Heisenberg algebras and $P_{\rm IV}$]{Polynomial Heisenberg algebras and $\boldsymbol{P_{\rm IV}}$}\label{section3} %\label{secPHA}

A polynomial Heisenberg algebra (PHA) is a particular deformation of the Heisenberg--Weyl algebra \cite{FH99, VS93}. Although there is a general def\/inition of a $m$-th order PHA, in this work we will only be interested in the second-order case.

A second-order PHA is def\/ined by{\samepage
\begin{gather*}
[H,L^{\pm}] =\pm 2 L^{\pm},\qquad
[L^{-},L^{+}] =Q(H+2)-Q(H)=P(H),
\end{gather*}
where $P$ and $Q$ are second- and third-order polynomials, respectively.}

\looseness=-1
This def\/inition means that $L^{\pm}$ act like ladder operators for the system described by the Hamiltonian~$H$~\cite{CFNN04,FNN04}. As usual, $L^{+}$ will be a creation operator and $L^{-}$ an annihilation one,~i.e.,
\begin{gather*}
L^{\pm}\psi_{k}\propto \psi_{k\pm 1}.
\end{gather*}

\begin{figure}[t]\centering
\includegraphics[scale=0.48]{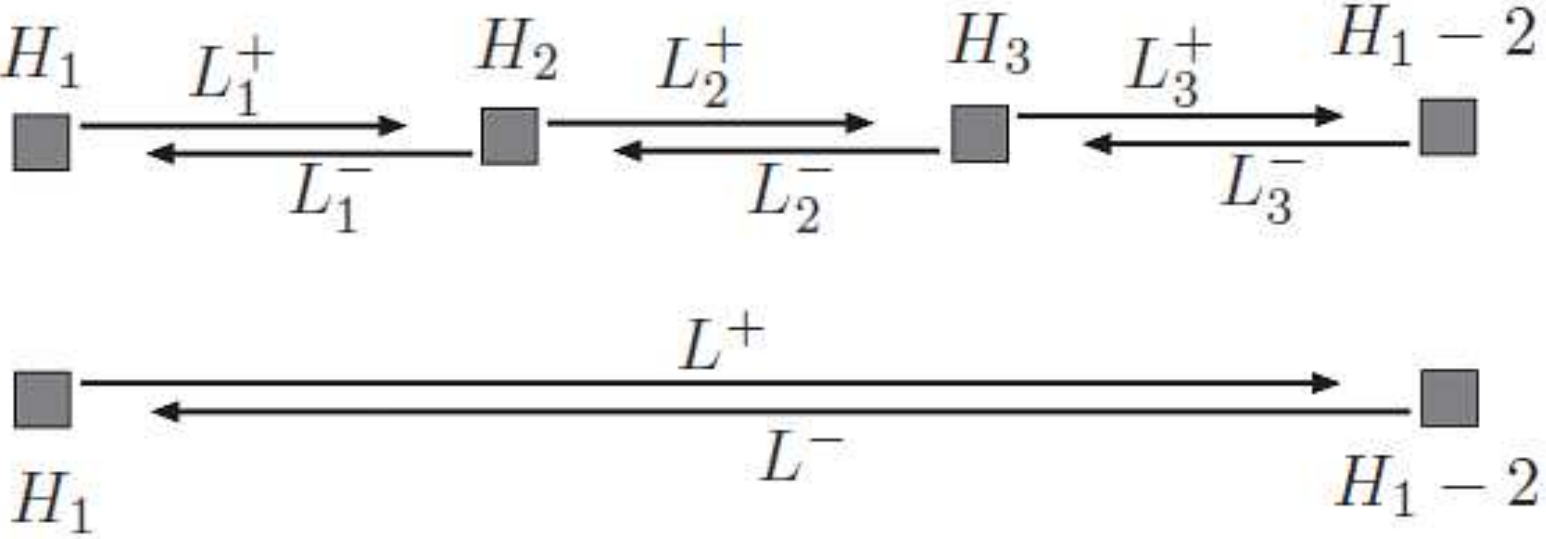}

\caption{Diagram of the two equivalent SUSY transformations. Above: the three-step f\/irst-order operators $L_1^{\pm}$, $L_2^{\pm}$, and $L_3^{\pm}$ allow to accomplish the transformation. Below: the direct transformation achieved through the third-order operators~$L^{\pm}$.}\label{diasusy}
\end{figure}

In this case, both $L^{\pm}$ will be third-order dif\/ferential ladder operators. Now, we propose a~closed-chain of three SUSY transformations \cite{Adl94,ACIN00,DEK94,Gra04,Mar09, VS93} so that $L^{\pm}$ can be expressed~as
\begin{gather}
L^{+} =L_{3}^{+}L_{2}^{+}L_{1}^{+}=(\partial-f_3)(\partial-f_2)(\partial-f_1),\nonumber\\
L^{-} =L_{1}^{-}L_{2}^{-}L_{3}^{-}=(-\partial-f_1)(-\partial-f_2)(-\partial-f_3).\label{eles} %\label{lmenos}
\end{gather}
In general, $(L^{-})^{\dag}\neq L^{+}$, except in the case where all $f_i \in \mathbb{R}$. As a matter of fact, in this article we are interested precisely in the case where $f_i\not\in\mathbb{R}$. Each $L_i$ fulf\/ills two intertwining relationships of kind
\begin{gather}
H_{i+1}L^{+}_i =L^{+}_{i}H_{i},\qquad
H_{i}L^{-}_i =L^{-}_{i}H_{i+1},
 \label{fac2}
\end{gather}
where $i=1,2,3$. In Fig.~\ref{diasusy} we present a diagram of the SUSY transformations.

If we equate the two dif\/ferent factorizations associated with \eqref{fac2} which lead to the same Hamiltonians we get
\begin{gather*}
H_{i+1}=L_{i}^{+}L_i^{-}+\epsilon_i=L_{i+1}^{-}L_{i+1}^{+}+\epsilon_{i+1}, \qquad i=1,2.
\end{gather*}
In addition, the closure condition is given by
\begin{gather*}
H_4=L_3^{+}L_3^{-}+\epsilon_3=H_1-2=L_1^{-}L_1^{+}+\epsilon_1-2.
\end{gather*}
By making the corresponding operator products we get the following system %of equations
\cite{Adl94,MN08,VS93}
\begin{subequations}
\begin{gather}
f_1'+f_2' =f_1^2-f_2^2+\epsilon_1-\epsilon_2,\label{f1}\\
f_2'+f_3' =f_2^2-f_3^2+\epsilon_2-\epsilon_3,\label{f2}\\
f_1'+f_3' =f_3^2-f_1^2+\epsilon_3-\epsilon_1+2.\label{f3}
\end{gather}
\end{subequations}

Eliminating $f_2^2$ from \eqref{f1} and \eqref{f2} we get
\begin{gather*}
f_1'+2f_2'+f_3'=f_1^2-f_3^2+\epsilon_1-\epsilon_3,
\end{gather*}
and from here we substitute $f_1^2$ from \eqref{f3} to obtain
\begin{gather*}
f_1'+f_2'+f_3'=1,
\end{gather*}
which, after integration becomes
\begin{gather}
f_1+f_2+f_3=x.\label{f4}
\end{gather}
Now, substituting \eqref{f4} into \eqref{f1}
\begin{gather*}
f_1=\frac{x-f_3}{2}+\frac{1-f_3'}{2(x-f_3)}-\frac{\epsilon_1-\epsilon_2}{2(x-f_3)}.
\end{gather*}
Let us def\/ine now a useful new function as $g \equiv f_3-x$, from which we get
\begin{gather*}
f_1=-\frac{g}{2}+\frac{g'}{2g}+\frac{\epsilon_1-\epsilon_2}{2g}.%\label{f2b}
\end{gather*}
Similarly, by plugging \eqref{f4} into \eqref{f2} and using $g$ we obtain
\begin{gather*}
f_2=-\frac{g}{2}-\frac{g'}{2g}-\frac{\epsilon_1-\epsilon_2}{2g},%\label{f3b}
\end{gather*}
Now that we have $f_1$, $f_2$, $f_3$ in terms of $g$, we replace them in \eqref{f3} in order to obtain
\begin{gather*}
gg'' = \frac{1}{2}(g')^2 + \frac{3}{2}g^4 + 4g^3x+ 2g^2\left(x^2+\epsilon_1+1-\frac{\epsilon_1+\epsilon_2}{2}\right)-\frac{(\epsilon_1-\epsilon_2)^2}{2},
\end{gather*}
which is the Painlev\'e IV equation ($P_{\rm IV}$)
\begin{gather*}
gg'' = \frac{1}{2}(g')^2 + \frac{3}{2}g^4 + 4g^3x+ 2g^2\left(x^2-a\right)+b,
\end{gather*}
with parameters
\begin{gather*}
a=\frac{\epsilon_1+\epsilon_2}{2}-\epsilon_3 -1,\qquad b=-\frac{(\epsilon_1-\epsilon_2)^2}{2}.
\end{gather*}
Since, in general $f\in\mathbb{C}$ then $g\in\mathbb{C}$. In addition, $\epsilon_i\in\mathbb{C}$ which implies that $a,  b\in\mathbb{C}$ and so~$g$ is a complex solution to~$P_{\rm IV}$ associated with the complex parameters~$a$,~$b$.

\section{First-order SUSY partners of the harmonic oscillator}\label{section4}

In this section we will show that the complex f\/irst-order SUSY transformation applied to the harmonic oscillator leads to a system with third-order ladder operators which is naturally ruled by a second-order PHA.

Let us begin with the harmonic oscillator potential
\begin{gather*}
V(x)=x^2.
\end{gather*}
As in Section \ref{section2}, we propose two operators $A^{+}$ and $A^{-}$ that fulf\/ill equations \eqref{As} and \eqref{entre1} so that
\begin{gather*}
\tilde{V}=V-2\beta'(x),
\end{gather*}
with the condition that $\beta(x)$ should solve the Riccati equation
\begin{gather}
\beta'+\beta^2=V-\epsilon.\label{ric}
\end{gather}

The previous scheme can be written in terms of a solution of the initial stationary Schr\"odinger equation $u$ by substituting $\beta=(\ln\, u)'$ in~\eqref{ric} in order to obtain
\begin{gather*}
-u''+V u = \epsilon u,
\end{gather*}
whose general solution, for any $\epsilon\in\mathbb{C}$, is given by
\begin{gather}
u(x)=\exp(-x^2/2)\left[ {}_1F_1\left(\frac{1-\epsilon}{4},\frac{1}{2};x^2\right)+(\lambda+i\kappa)\,x\, {}_1F_1\left(\frac{3-\epsilon}{4},\frac{3}{2};x^2\right)\right].
\label{solu}
\end{gather}
With this formalism, the f\/irst-order SUSY partner potential $\tilde{V}$ of the harmonic oscillator is given by
\begin{gather*}
\tilde{V}(x)=x^2-2[\ln u(x)]''.
\end{gather*}
The previously known results for the real case~\cite{JR98} are obtained by taking $\epsilon\in\mathbb{R}$, $\kappa=0$ and expressing $\lambda$ as
\begin{gather*}
\lambda=2\nu\frac{\Gamma\left(\frac{3-\epsilon}{4}\right)}{\Gamma\left(\frac{1-\epsilon}{4}\right)}.
\end{gather*}
On the other hand, for $\epsilon\in\mathbb{C}$ the transformation function $u(x)$ is complex and so is $\tilde{V}(x)$.

\begin{figure}[t]
\centering
\includegraphics[scale=0.43]{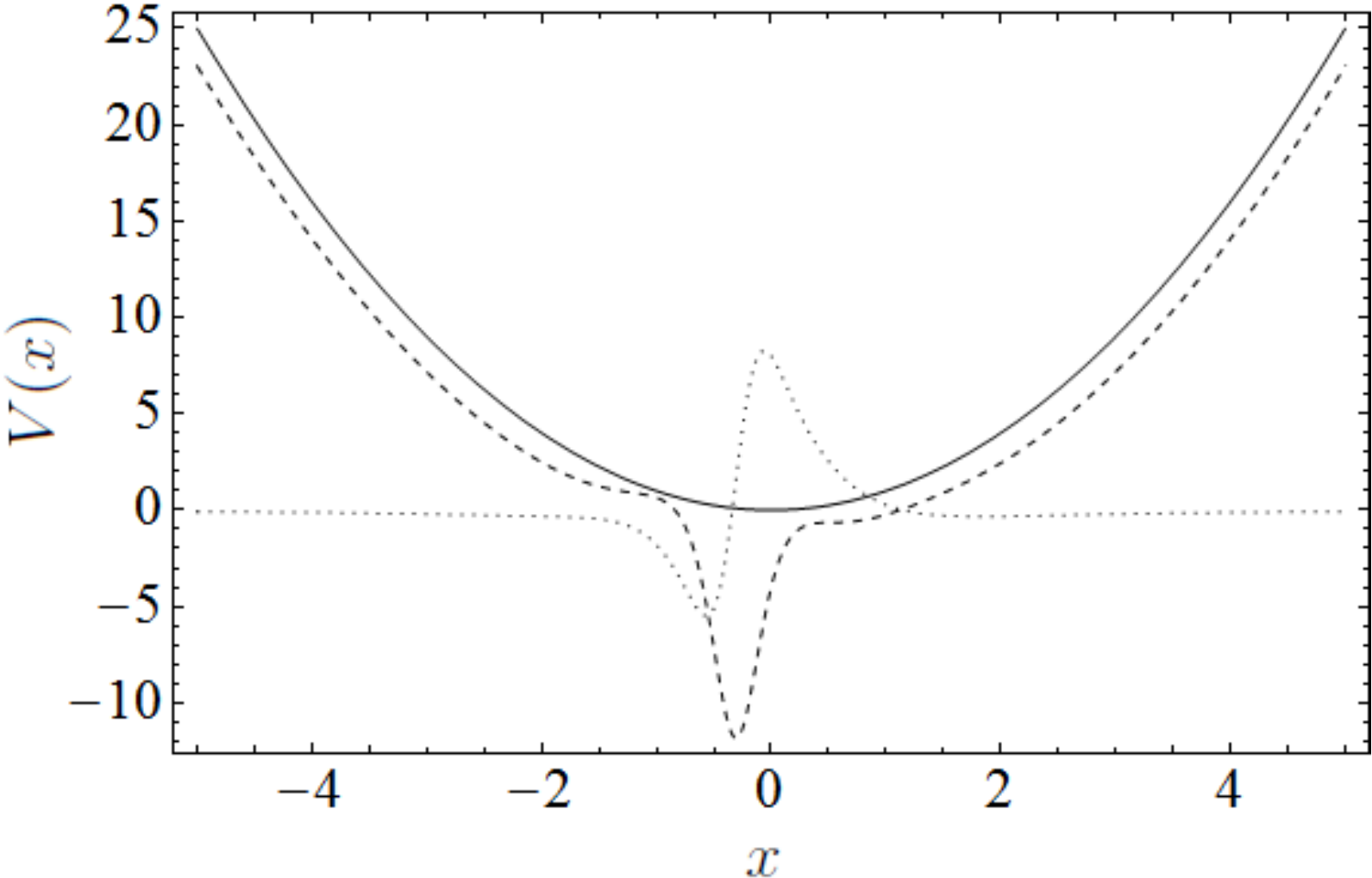} \hspace{8mm}
\includegraphics[scale=0.43]{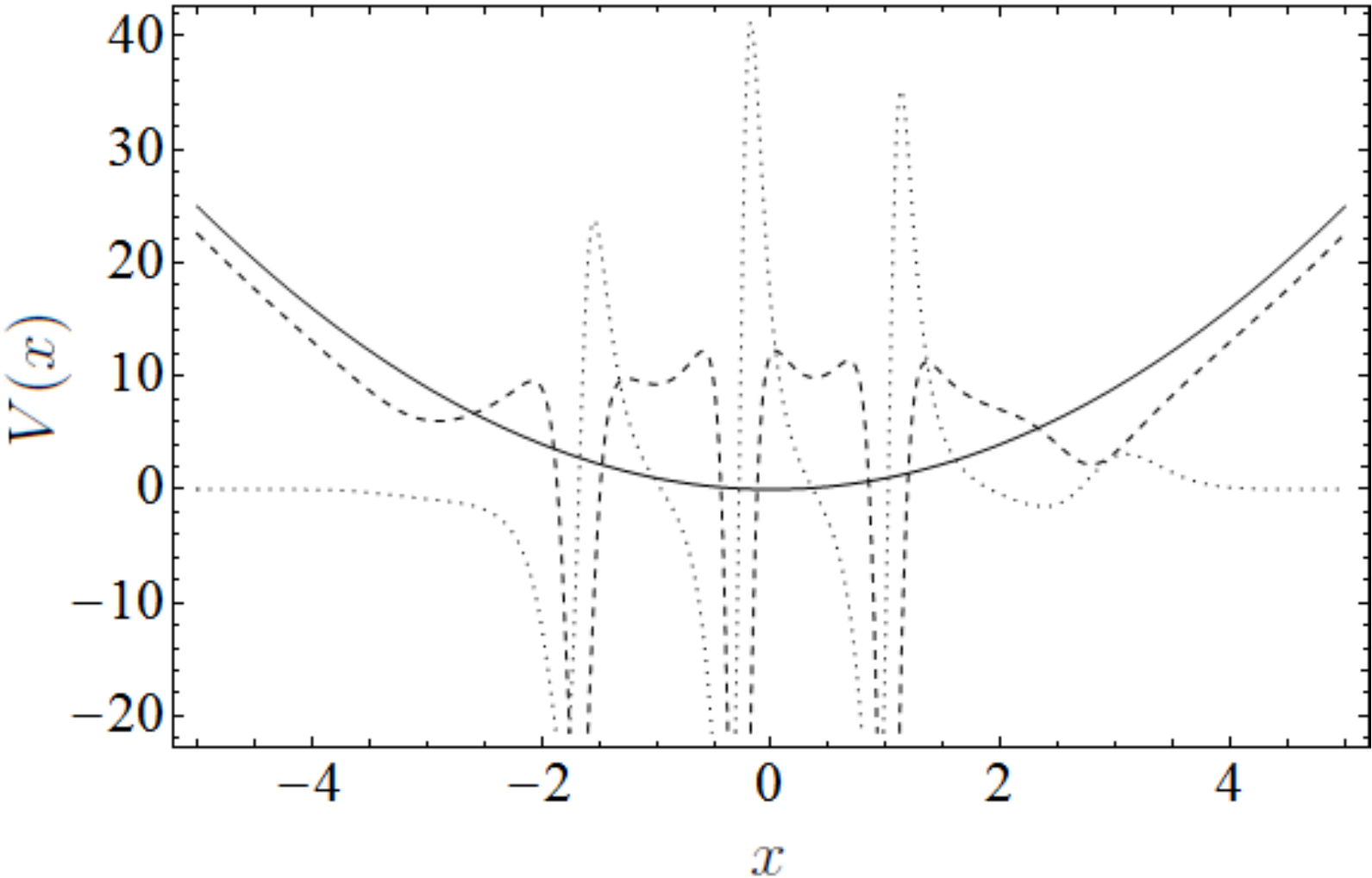}

\caption{Examples of SUSY partner potentials of the harmonic oscillator using the two complex factorization energies $\epsilon=-1+i$ with $\lambda=\kappa=1$ (left) and $\epsilon=3+i10^{-3}$ with $\lambda=\kappa=2$ (right). Its real (dashed line) and imaginary (dotted line) parts are compared to the harmonic oscillator (solid line).}\label{susypot}
\end{figure}

\begin{figure}[t]
\centering
\includegraphics[scale=0.34]{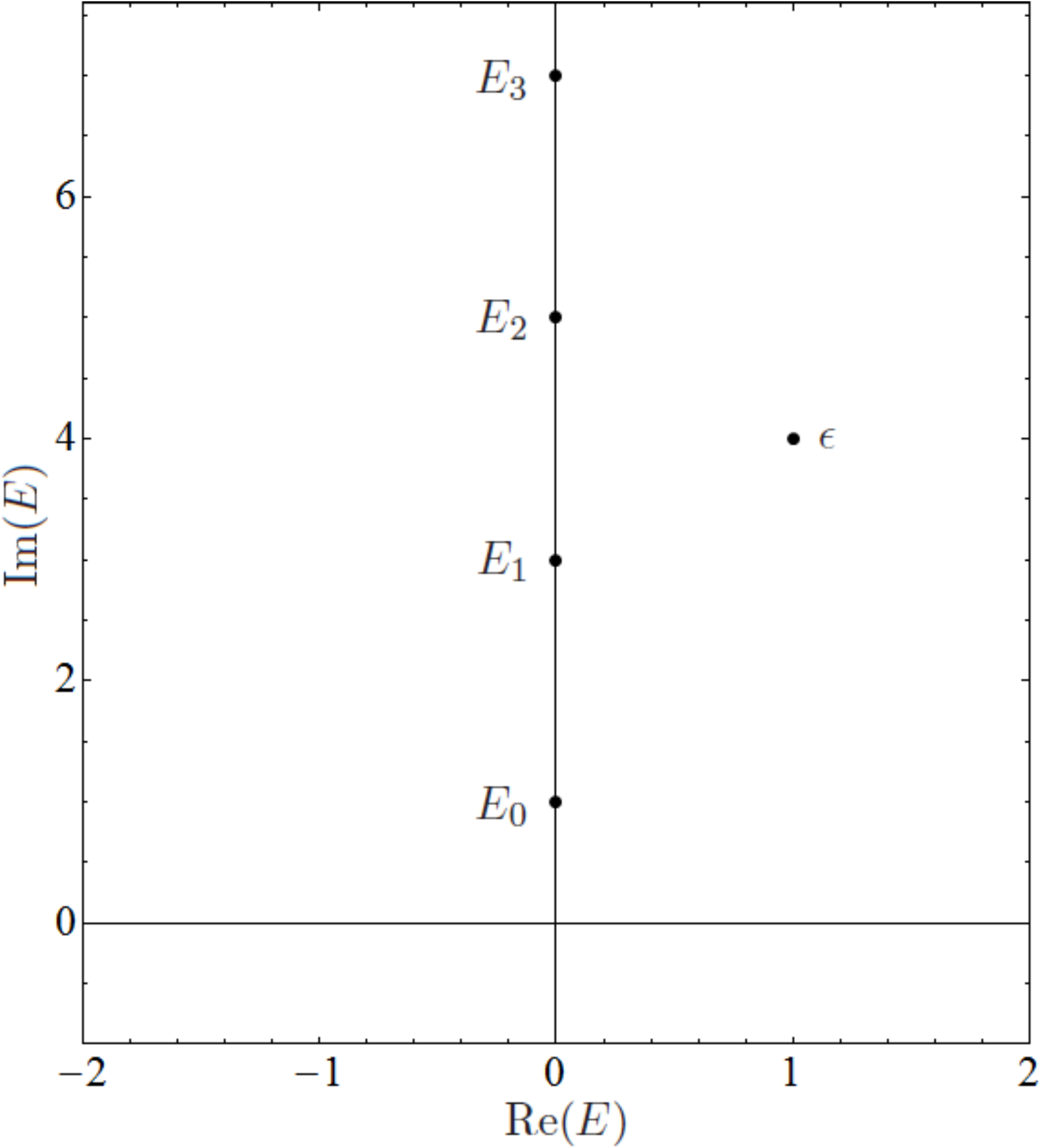}

\caption{The complex energy plane which contains the eigenvalues of the SUSY generated Hamiltonian~$\tilde{H}$. Along the real line it is drawn the usual ladder formed by the energy levels $E_n=2n+1$, $n= 0,1,\dots $, and out of the real line there is a new level of the complex value~$\epsilon$.}
\label{parspace}
\end{figure}

In Fig.~\ref{susypot} we present some examples of complex SUSY partner potentials of the harmonic oscillator generated for $\epsilon\in\mathbb{C}$ and compare them with the initial potential. Let us note that these new potentials have the same real spectra as the harmonic oscillator, except that they have one extra energy level, located at the complex value $\epsilon$. This kind of spectrum is represented in the diagram of Fig.~\ref{parspace}. Note that this can be interpreted as the superposition of the two ladders shown in equation~\eqref{spectrum}.

\section{Complex solutions to Painlev\'e IV equation}\label{section5}

In this Section we will see that for each f\/irst-order SUSY partner of the harmonic oscillator we will have three dif\/ferent complex solutions of $P_{\rm IV}$.

First of all, let us note that the ladder operators associated with $\tilde{H}$ are given by
\begin{gather*}
L^{\pm}=A^{+}a^{\pm}A^{-},%\label{lmas2}
\end{gather*}
or, explicitly,
\begin{gather}
L^{+} =(-\partial +\beta)(-\partial+x)(\partial+\beta)=(\partial -\beta)(\partial-x)(\partial+\beta),\nonumber\\
L^{-} =(-\partial +\beta)(\partial+x)(\partial+\beta)=(-\partial +\beta)(-\partial-x)(-\partial-\beta).\label{eles2}
\end{gather}
This system has third-order dif\/ferential ladder operators, therefore it is ruled by a second-order PHA and so we can apply our analysis of Section~\ref{section3} by identifying the ladder operators $L^{\pm}$ given in equations~\eqref{eles2} with those of equations~\eqref{eles}. This identif\/ication leads to
\begin{gather}\label{efes}
f_1= -\beta,\qquad
f_2= x,\qquad
f_3= \beta.
\end{gather}
Recall that the function $g=f_3-x$ fulf\/ills the Painlev\'e IV equation, then one solution to $P_{\rm IV}$ is
\begin{gather*}
g=\beta-x.
\end{gather*}

Let $\psi_{\mathcal{E}_i}$, $i=1,2,3$, be the states annihilated by $L^{-}$, where $\mathcal{E}_i$ represents the factorization energy for the corresponding extremal state. In particular, let $\psi_{\mathcal{E}_3}$ be annihilated by $L_3^{-}$, and from equation~\eqref{eles} we can see that it is also annihilated by~$L^{-}$. By solving $L_3^{-}\psi_{\mathcal{E}_3}=0$ we get
\begin{gather*}
\psi_{\mathcal{E}_3}\propto \exp\left[-\int f_3(y)dy\right],
\end{gather*}
from which it can be shown that the corresponding solution to $P_{\rm IV}$ reads
\begin{gather*}
g_3(x,\epsilon)=-x-{\ln[\psi_{\mathcal{E}_3}(x)]}'.
\end{gather*}

\begin{figure}[t]
\centering
\includegraphics[scale=0.42]{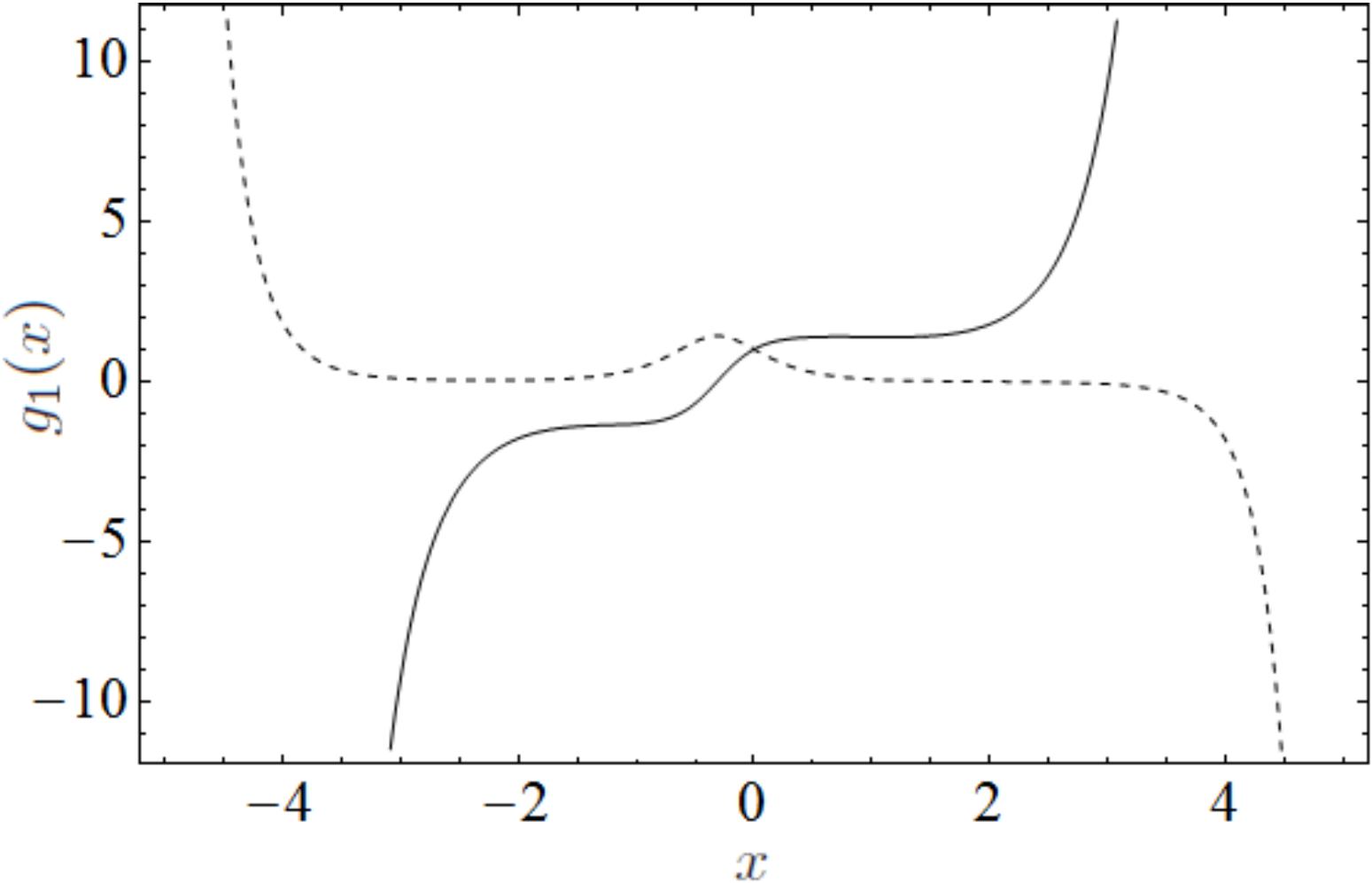}
\hspace{8mm}
\includegraphics[scale=0.42]{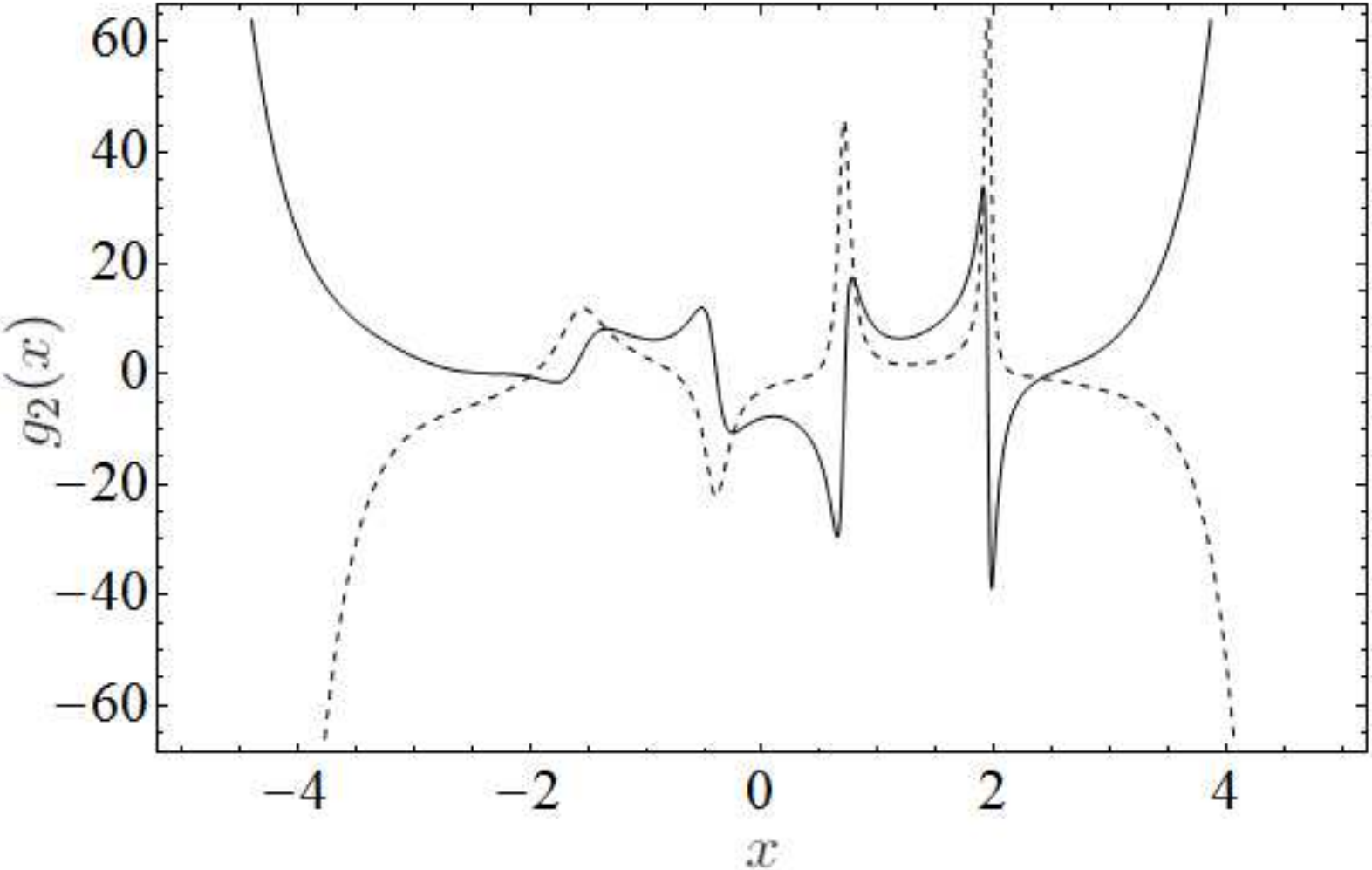}
\includegraphics[scale=0.42]{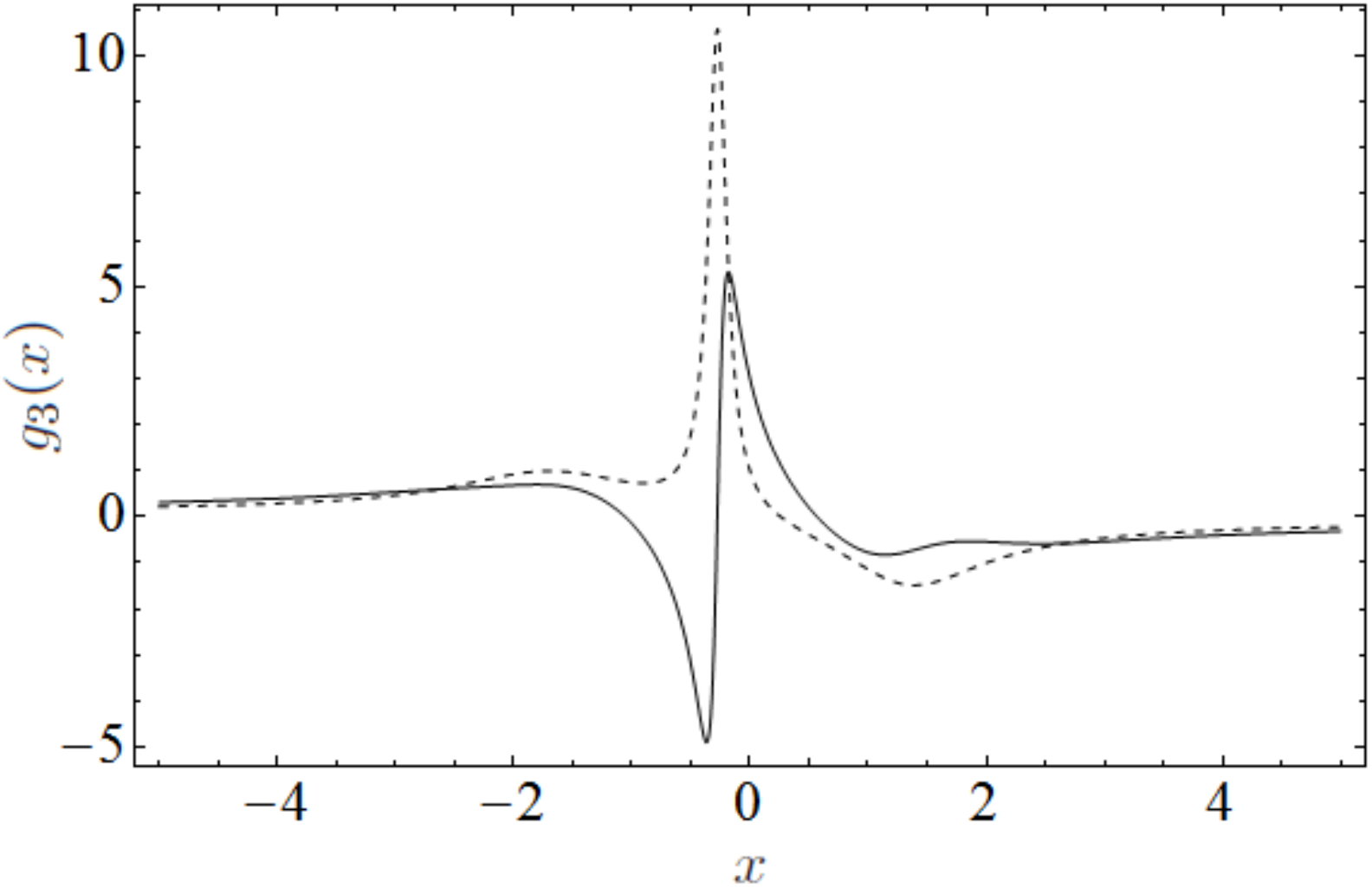}

\vspace{-1mm}

\caption{Complex solutions to $P_{\rm IV}$. The solid line corresponds to the real part and the dashed to the complex one. $g_1(x)$ for $\epsilon=-1+i10^{-2}$ and $\lambda=\kappa=1$; $g_2(x)$ for $\epsilon=4+i 2^{-1}$, $\lambda=\kappa=1$; and $g_3(x)$ is given for $\epsilon=1+i$ and $\lambda=3$, $\kappa=1$.}\label{sols}\vspace{-1mm}
\end{figure}

For our complex f\/irst-order SUSY partner potential $\tilde{V}(x)$, generated by using the complex seed solution $u(x)$ of equation~\eqref{solu}, the three extremal states (up to a numerical factor) and their corresponding energies consistent with equations~\eqref{efes} are given by:
\begin{alignat}{3}
& \psi_{\mathcal{E}_1} = A^{+}a^{+}u, \qquad && \mathcal{E}_1 = \epsilon+2
, \qquad
  \psi_{\mathcal{E}_2} = A^{+}\exp\big({-}x^2/2\big), \qquad   \mathcal{E}_2 = 1, & \nonumber\\
&\psi_{\mathcal{E}_3} = u^{-1}, \qquad && \mathcal{E}_3 =  \epsilon.& \label{psis}
\end{alignat}
Nevertheless, the labelling given in equations~\eqref{psis} is not essential, so that by making cyclic per\-mu\-tations, associated with the three extremal states of equations~\eqref{psis} we get three solutions to $P_{\rm IV}$:
\begin{gather*}
g_i(x,\epsilon)=-x-{\ln[\psi_{\mathcal{E}_i}(x)]}',
\end{gather*}
with $i=1,2,3$. The corresponding parameters $a$, $b$ of $P_{\rm IV}$ are given by:
\begin{alignat}{3}
& a_1=-\frac{1}{2}(\epsilon+5), \qquad && b_1=  -\frac{1}{2}\left(\epsilon-1\right)^2,
\qquad
  a_2=  \epsilon-1,\qquad   b_2= -2, & \nonumber\\
&a_3=  \frac{1}{2}(1-\epsilon), \qquad && b_3=  -\frac{1}{2}\left(\epsilon+1\right)^2, &\label{abs}
\end{alignat}
where we have added the subscript corresponding to the extremal state used. In Fig.~\ref{sols} we have presented one example for each of the three families of solutions.

\begin{figure}[t]
\centering
\includegraphics[scale=0.33]{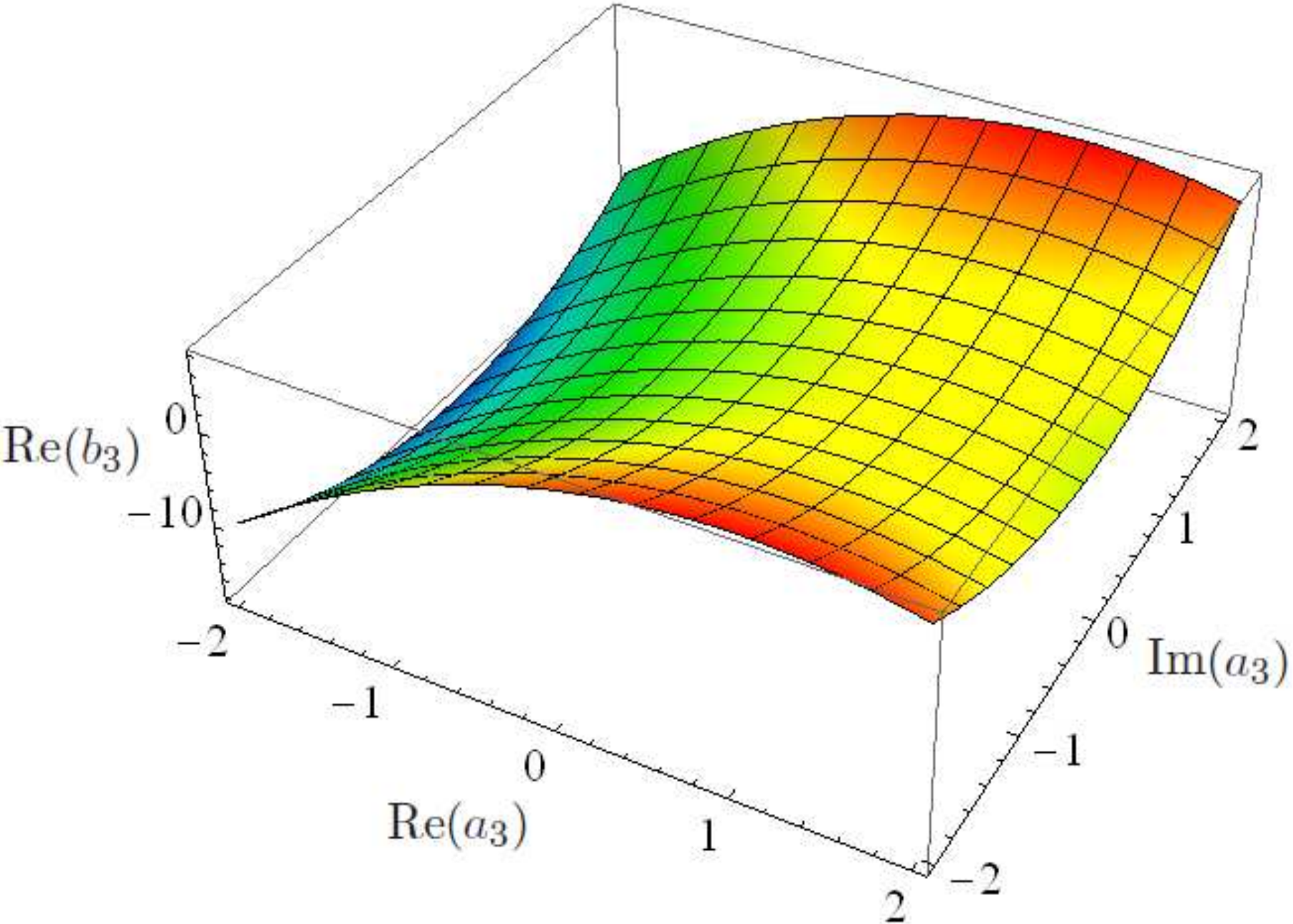}
\hspace{8mm}
\includegraphics[scale=0.33]{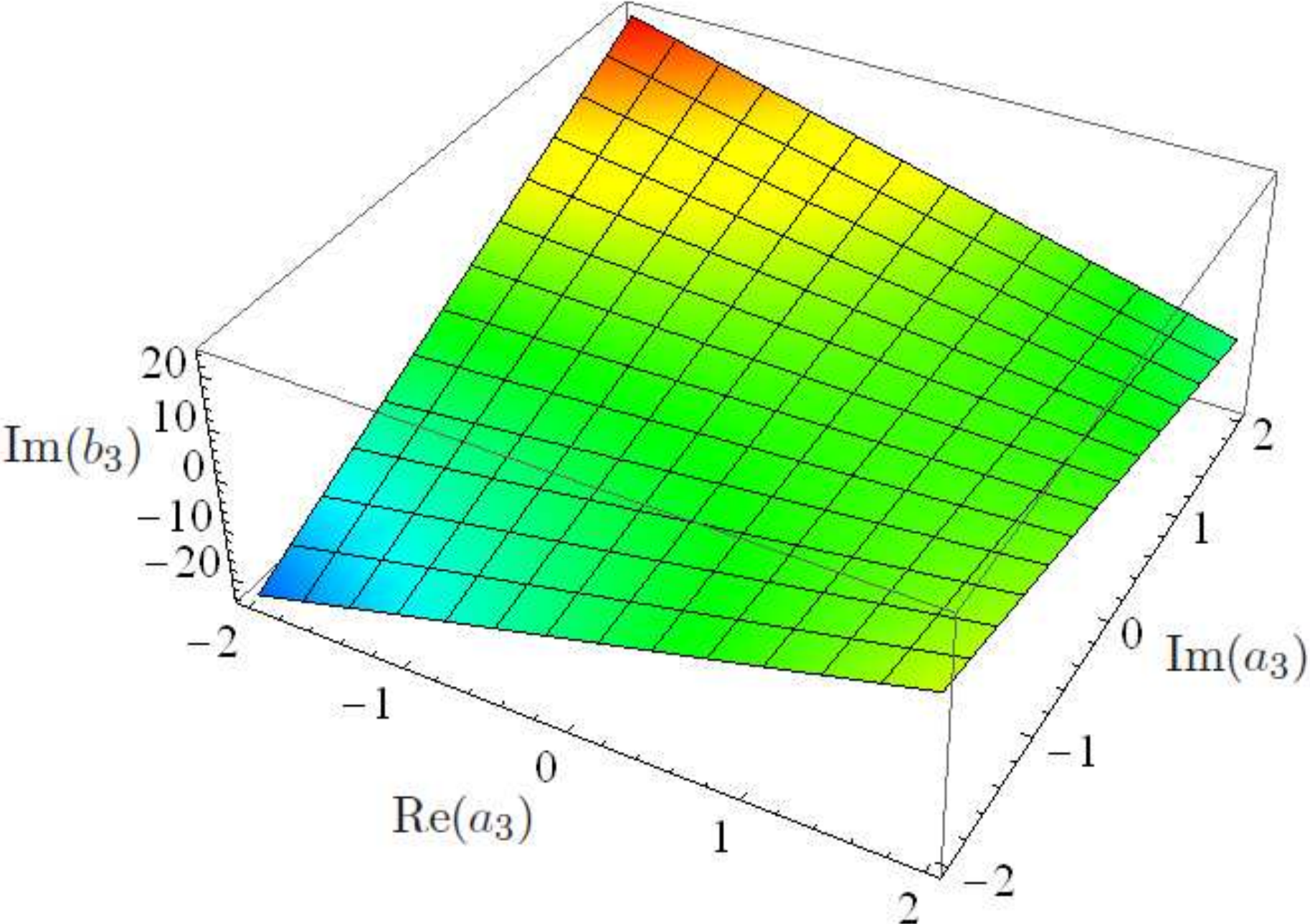}

\vspace{-1mm}

\caption{Parameter space where we show $\text{Re}(b_3)$ (left) and $\text{Im}(b_3)$ (right) in function of $\text{Re}(a_3)$ and $\text{Im}(a_3)$. The plots for $b_1 (a_1)$ and $b_2 (a_2)$ are similar.}\label{figpara}
\vspace{-2mm}
\end{figure}

From equations~\eqref{abs} we can see that $a_i$ is linear in $\epsilon$ for all three cases. So, instead of studying the parametric relationship of $a,b$ in terms of $\epsilon$ let us analyze $b_i=b_i(a_i)$, namely,
\begin{gather}
b_1=  -2(a_1+3)^2,\qquad b_2= -2,\qquad
b_3=  -2(a_3-1)^2.\label{b1a1}
\end{gather}
Then, we can choose $a_i\in\mathbb{C}$ but $b_i$ will be f\/ixed by its corresponding relationship with $a_i$. In Fig.~\ref{figpara} we show the domain for $b_3$ from equations~\eqref{b1a1}. There are similar plots for $b_1$ but for $b_2$ remember that $b_2=-2$ $\forall\, a_2\in\mathbb{C}$.

\section{Conclusions}\label{section6}

In this work we have applied the f\/irst-order SUSY QM to the harmonic oscillator potential to obtain systems with third-order ladder operators which are ruled by second-order PHA. From this system we have obtained complex solutions $g_1$, $g_2$, and $g_3$ of the Painlev\'e IV equation.

\looseness=-1
It is worth to notice that complex solutions have not been generated in this context, nevertheless, the work of Bassom et al.~\cite[Section~3.3]{BCH95} is general and can be applied to a $P_{\rm IV}$ with complex parameters. An extension of the parameter space to the complex plane is helpful at least when~$P_{\rm IV}$ appears in the reduction of other non-linear dif\/ferential equations. Moreover, it can be used in the future to study complex extensions of those physical systems described by~$P_{\rm IV}$.

We f\/irst discussed the complex f\/irst-order SUSY transformation for general systems. We def\/ined then the PHA and established explicitly the relationship of the second-order case with the Painlev\'e IV equation. We also studied the complex SUSY partner potentials of the harmonic oscillator with an analysis of the energy spectra of the new potentials. Then, we obtained complex solutions of the Painlev\'e IV equation and we analized the complex parameter space where the solutions can appear. We concluded that there are three dif\/ferent families of complex solutions for each complex f\/irst-order SUSY transformation.

In the future we would like to address the higher-order SUSY transformation in order to extend the space of complex solutions of the Painlev\'e IV equation and to study in more detail the properties of these solutions.

\vspace{-3mm}

\subsection*{Acknowledgements}

The author would like to acknowledge the useful comments of Professor David J.~Fern\'andez~C. and his revision of the manuscript. The author also acknowledges the support of Conacyt (Mexico) through the PhD scholarship 219665 and the project 152574.

\vspace{-3mm}

\pdfbookmark[1]{References}{ref}
\LastPageEnding

\end{document}